\documentclass[12pt]{iopart}
\usepackage{xcolor}

\usepackage{lineno}
\usepackage{wasysym,amssymb}
\usepackage{natbib}
\usepackage{graphicx}

\newcommand{\upd}{\mathrm{\,d}}

\newcommand{\red}[1]{\textcolor{black}{#1}}

\begin{document}

 % \linenumbers
\title{Multi-level segment analysis: definition and application in turbulent systems}

\author{L.P. Wang$^1$, Y.X. Huang$^2$}
\address{$^{1}$UM-SJTU Joint Institute, Shanghai JiaoTong University, Shanghai, 200240, China\\$^2$State Key Laboratory of Marine Environmental Science,
Xiamen University, Xiamen 361102, PR China}
\ead{lipo.wang@sjtu.edu.cn,yongxianghuang@gmail.com}

\begin{abstract}
For many complex systems the interaction of different scales is among the most interesting and
challenging features. It seems not very successful to extract the physical properties in different scale regimes by the existing approaches, such as structure-function and Fourier spectrum method. Fundamentally these methods have their respective limitations, for instance scale mixing, i.e. the so-called infrared and ultraviolet effects. To make improvement in this regard, a new method, multi-level segment analysis
(MSA) based on the local extrema statistics, has been developed. Benchmark (fractional Brownian motion) verifications and the important case tests (Lagrangian and two-dimensional turbulence) show that MSA can successfully reveal different scaling regimes, which has been remaining quite controversial in turbulence research. In general the MSA method proposed here can be applied to different dynamic systems in which the concepts of multiscaling and multifractal are relevant.
\end{abstract}

%\pacs{94.05.Lk, 47.27.Gs,05.45.Tp}%{Time series analysis}
\maketitle
\section{Introduction}

Multiscale is one of the most important features commonly existing in complex systems, where a large range of spatial/temporal scales coexist and interact with each other. Typically such interaction generates scaling relations in the respective scale ranges. To understand the multiscale statistical behavior has been remaining as the research focus in various areas, such as fluid turbulence~\citep{Frisch1995Book}, financial market analysis \citep{Mantegna1996Nature,Schmitt1999,Li2014PhysicaA}, environmental science~\citep{Schmitt1998JGR} and population dynamics \citep{Seuront1999JPR}, to list a few. Among a number of existing analysis approaches for such problems, the standard and most cited one is structure-function (SF), which is first introduced by Kolmogorov in his famous homogeneous and isotropic turbulence theory in 1941~\citep{Kolmogorov1941,Frisch1995Book}. However, the average operation in SF mixes regions with different correlations~\citep{Wang2006JFM,Wang2008JFM}. Mathematically, SF acts as a filter with a weight function of $W(k,\ell)=1-\cos(2\pi k \ell)$, in which $k$ is the wavenumber and $\ell$ is the separation scale \citep{Davidson2005PRL,Huang2010PRE}. It thus makes the statistics at different scale $\ell$ strongly mixed, resulting in the so-called infrared and ultraviolet effects, respectively for large-scale and small-scale contamination~\citep{Huang2013PRE}. The situation will be more serious when an energetic structure presents, e.g., annual cycle in collected geoscience data \citep{Huang2009Hydrol}, large-scale circulations in Rayleigh-B\'{e}nard convection, vortex trapping events in Lagrangian turbulence \citep{Huang2013PRE,Wang2014PoF}.

Fourier analysis in the frequency domain has the similar deficiency as SF, i.e. any local event will propagate the influence over the entire analyzed domain, especially for the nonlinear and nonstationary turbulent structures~\citep{Huang1998EMD,Huang2011PRE}. \citet{Farge1992ARFM} claimed that a localized expansion should be preferred over unbounded trigonometric functions used in Fourier analysis, because it is believed that trigonometric functions are at risk of misinterpreting the characters of field phenomena. An alternative approach, namely wavelet transform is then proposed to overcome the possible shortcoming of the Fourier transform with local capability~\citep{Daubechies1992,Farge1996IEEE}. \red{However, the same problem as Fourier analysis still exists, if the fixed mother wavelet has a shape different from the analyzed data structure. We also note that the classical structure-function analysis is referred to as `the poor man's wavelet'\citep{Lovejoy2012NPG}.}

% As pointed by~\citet{Huang1998EMD}, wavelet can be considered as a windowed %Fourier analysis. It thus inherits more or less the same deficiency as the %Fourier analysis since the basis function, namely mother wavelet is proposed %\textit{a priori}, which can cause the high-order harmonic problem \citep{Cohen1995Book,Flandrin1998Book,Huang2011PRE}.
%The scaling behavior extracted by Fourier or Wavelet analysis therefore %may be contaminated by nonlinear and nonstationary events, such as the ramp-cliff %structure in passive scalar turbulence \citep{Warhaft2000ARFM,Huang2011PRE}.

To overcome the potential weaknesses of SF or Fourier analysis, several
methodologies have been proposed in recent years to emphasize the local geometrical features, such as \red{wavelet-based methodologies (wavelet leader \citep{Jaffard2005wavelet,Lashermes2008EPJB}, wavelet transform modulus maxima \citep{Muzy1993PRE,Oswicecimka2006PRE}, etc.),} detrended fluctuation analysis \citep{Peng1994PRE}, detrended structure-function \citep{Huang2014JoT}, scaling of maximum probability density function of increments \citep{Huang2011PoF,Qiu2014JHydrol}, and Hilbert spectral analysis \citep{Huang2008EPL,Huang2011PRE}, to name a few. Note that different approaches may have different performances, and their own advantages and disadvantages. For example, the detrended structure-function can constrain the influence of the large-scale structure, using the detrending procedure to remove the scales larger than the separation distance $\ell$. In practice, the famous $4/5$-law can then be more clearly retrieved than the classical SF \citep{Huang2014JoT}. However, this method is still biased with the vortex trapping event in Lagrangian turbulence, which typically possesses a time scale around $3\sim 5\tau_{\eta}$ in the dissipative range \citep{Toschi2005JOT}. The scaling of maximum probability density function of increments helps to quantify the background fluctuation of turbulent fields. Compared with SF, it can efficiently extract the first-order scaling relations \citep{Huang2011PoF,Qiu2014JHydrol}; however, it is difficult to extend to higher $q$th-order cases.

A new view on the field structure is based on the topological features of the extremal points. In principle, physical systems may assume different complexity and interpretability in different spaces, such as physical or Fourier~\citep{Wang2013PTRS}. The extremal point structure in physical space has the straightforward advantage in defining characteristic parameters. Considering a fluctuating quantity, turbulence disturbs the flow field to generate the local extrema, while viscous diffusion will smooth the field to annihilate the extremal points. By nature the statistics of local extremal points inherits the process physics. Based on this idea, Wang, Peters and other collaborators have studied passive scalar turbulence via dissipation element analysis~\citep{Wang2006JFM,Wang2008JFM}. \citet{Wang2014PoF} analyzed the Lagrangian velocity by defining the trajectory segment structure from the extrema of particles' local acceleration. Such diagnosis verifies successfully the Kolmogorov scaling relation, which has been argued controversially for a long time \citep{Falkovich2012PoF}. However, under some circumstances the extremal points may largely be contaminated by noise, thus partly be spurious. In other words extremal points are sensitive to noise perturbation. Although data smoothing can relieve this problem, some artificial arbitrariness will inevitably be introduced; moreover it may not be easy to design reliable smoothing algorithms from case to case. In this regard the extrema-based analyses are not generally applicable, e.g. with noises from measurement inaccuracy, interpolation error or external perturbations.

In this paper a new method, multi-level segment analysis (MSA), has been developed. The key idea hereof is based on the observation that local extrema are conditionally valid, indicating a kind of multi-level structure. Compared with the aforementioned extrema-based analyses, this new  method is a reasonable extension with more applicability. Details in algorithm definition, verification and applications will be introduced in the following.

%\section{Methodology}
\section{Multi-level segment analysis: method definition}
\begin{figure}
\centering
\includegraphics[width=0.65\linewidth,clip]{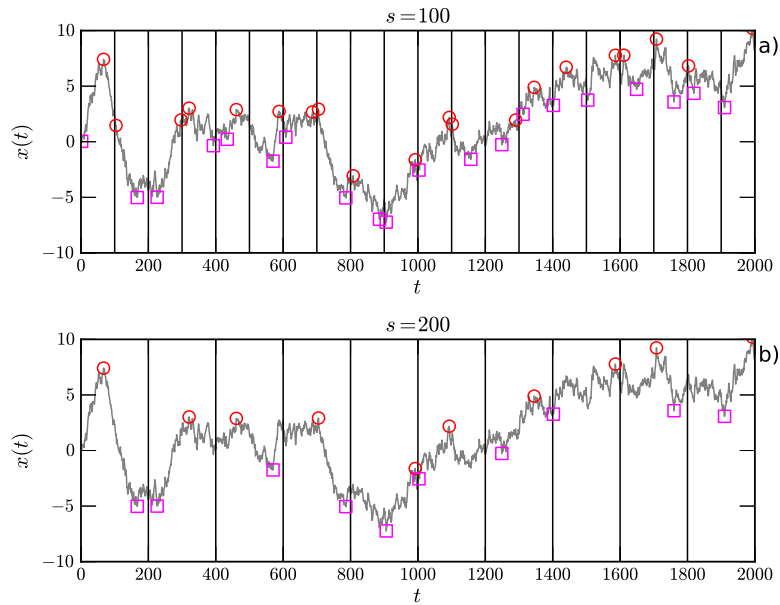}
\caption{(Color online) \red{An illustration of the extracted} extreme points at different $s$ levels: a) $s=100$, and b) $s=200$. The local maxima and minima are demonstrated respectively by $\ocircle$ and $\square$. The vertical line indicates the window size $s$. \red{To ensure spacial homogeneity, a sliding window with level $s$ scans over the whole data set, see more detail in the text.}} \label{level}
\end{figure}

Considering any function $f(x)$ in some physical process, where $x$ is the
independent variable, e.g., the spatial or temporal coordinates, $x_0$ is a local
extremal point with respect to scale $s$ is defined as $f(x_0) \leq f(x), \forall x\in
(x_0-s,x_0+s)$ (minimum), or $f(x_0)\ge f(x), \forall x\in (x_0-s,x_0+s)$
(maximum). Extrema are conditionally valid. For instance, if $x_0$ is extremal at
scale $s$, it may not be extremal at a larger scale $s_{1}>s$. Figure~\ref{level}
illustrates that for an artificially generated signal, at different $s$ levels both the
number density and the fluctuation amplitude of the extremal points will change accordingly. Under some special conditions extremal points have simple but interesting properties. For instance, for a monotonous function, there is no extrema for any $s$; for a single harmonic wave function with a period of $T$, the number of extremal point remains constant. For real complex multiscale systems as turbulence, variation of extremal point is continuously dependant on $s$.

At a specific $s$ level, denote the corresponding extremal point set as $x_{s,i}$, $i=1,2,...$ (along the coordinate increasing direction). Numerical tests show that typically these points are $\max-\min$ alternated for small $s$, while when $s$ increases $\max-\max$ or $\min-\min$ events may also appear but with low probability (e.g., $5\%$), depending on the $f(x)$ structure and the process physics. The segment is defined as the part of $f(x)$ between two adjacent extremal points. The characteristic parameters to describe the structure skeleton are the function difference, i.e., $f(x_{s,i})-f(x_{s,i-1})$ and the length scale, i.e., $\ell=x_{s,i}-x_{s,i-1}$. By varying the $s$ value, different extremal point sets, and thus different segment sets, can be obtained. In this sense this procedure is named as multi-level segment analysis (MSA); while the existing approaches based on local extremal points~\citep{Wang2006JFM,Wang2008JFM,Wang2014PoF} calculate extrema from the DNS data; thus can be understood as single-level. Compared with the conventional SF, in MSA the segment length scale is not an independent input, but determined by $f(x)$.

\red{For a specified $s$, scan over the data domain to have the corresponding segment characteristics, i.e. $f(x_{s,i})-f(x_{s,i-1})$ and $\ell=x_{s,i}-x_{s,i-1}$. Collect the results for different $s$, the functional statistics can be defined. Numerically the same segments may be counted repeatedly for different $s$, which then need to be excluded}. In terms of structure function, the mathematical expression is (for the $q$th order case)
\begin{equation}
\mathcal{D}_q(\ell)=
%\langle|\Delta \bar{\theta}|^{q}\rangle \equiv \sum_{ s}\sum_{i=1}^{n(s)}
\langle [f(x_{s,i})-f(x_{s,i-1})]^{q}\vert {x_{s,i}-x_{s,i-1}=\ell}\rangle_{s},
\label{newsf}
\end{equation}
 where $\langle\cdot\rangle_{s}$ denotes sampling over different $s$. If any scaling relation exists, one may expect a power-law behavior as
\begin{equation}
\mathcal{D}_q(\ell)\sim \ell^{\xi(q)},
\end{equation}
in which $\xi(q)$ is the MSA scaling exponent.

\red{For comparison we also include here briefly the classical SF and wavelet-leader definitions. The $q${th}
order SF is written as
\begin{equation}
  S_q(\ell)=\langle \vert \Delta x_{\ell}(t) \vert^q \rangle \sim
\ell^{\zeta(q)},
\end{equation}
where  $\Delta x_{\ell}(t)=x(t+\ell)-x(t)$ and $\ell$ is the time separation scale. The scaling exponent $\zeta(q)$ characterizes the fluctuation statistics. $\zeta(q)$ is linear for monofractal processes such as fractional Brownian motion, and nonlinear and concave for multifractal
processes \citep{Schertzer1997}. As mentioned above, SF mixes information from different scales. It is also limited by the slope of the Fourier spectrum $E(f)\sim f^{-\beta}$, e.g., $1<\beta<3$ \citep{Frisch1995Book,Huang2010PRE}. We denote this as $\beta$-limitation.}

\red{There are several different wavelet-based methods, for example, wavelet-transform-modulus-maxima (WTMM) \citep{Muzy1991PRL,Mallat1992IEEE,Muzy1993PRE} and wavelet-leader (WL) \citep{Jaffard2005wavelet,Wendt2007,Lashermes2008EPJB}. We consider here only WL. More detailed discussions of those methods can refer to Ref. \citep{Jaffard2005wavelet,Oswicecimka2006PRE,Huang2011PRE} and references therein.
\\The discrete wavelet transform is defined as
\begin{equation}
\psi(k,j)=\int_{\mathbb{R}}x(t)\varphi\left(  2^{-j}t-k  \right)\upd t,
\end{equation}
where $\varphi$ is the chosen wavelet, $\psi(k,j)$ is the wavelet coefficient, $k$ is the position index, $j$ is the scale index, and $\ell=2^j$ is the
corresponding scale \citep{Daubechies1992,Mallat1999Book}. Every discrete wavelet coefficient $\psi(k,j)$ can be associated with the dyadic interval $\varrho(k,j)$
 \begin{equation}
 \varrho(k,j)=[2^jk,2^j(k+1)).
 \end{equation}
Thus the wavelet coefficients can be represented as $\psi(\varrho)=\psi(k,j)$. The wavelet-leader is defined as
\begin{equation}
l(k,j)=\sup_{\varrho'\subset 3 \varrho(k,j),j'\le j} \vert \psi(\varrho')\vert,
\end{equation}
where $3 \varrho(k,j)=\varrho(k-1,j)\cup \varrho(k,j)\cup \varrho(k+1,j)$ \citep{Jaffard2005wavelet,Lashermes2008EPJB,Wendt2007}. The expected scaling behavior can be expressed as
\begin{equation}
Z_q(j)= \langle l(k,j)^q \rangle\sim 2^{j\chi(q)},
\end{equation}
in which $\chi(q)$ is the corresponding scaling exponent. The calculation efficiency has been discussed for various datasets \citep{Jaffard2005wavelet,Wendt2007,
 Lashermes2008EPJB}.}

As shown in the rest of this paper, for simple cases, MSA and the classic methods show pretty identical results; while for complex analyses as turbulence, because of the algorithmic principle to depict the function structure, the former one is more effective and efficient.

Some additional comments are stated as follows. First, MSA can be considered as a dynamic-based approach without any basis assumption \textit{a priori}. Since local extrema imply the change of the sign of $f'(x)$ across $x_{s,i}$. Assuming that $f(x)$ represents a velocity signal, $f'(x)$ is then the acceleration, i.e. a dynamical variable. Second, MSA shares the spirit of the wavelet-transform-modulus-maxima (WTMM) \citep{Muzy1993PRE}, in which only the maximum modulus of the wavelet coefficient is considered.  However, as argued in several Refs.\,\citep{Huang1998EMD,Huang2011PRE},
if the shape of the chosen mother wavelet is different with the specific turbulent structure, e.g., ramp-cliff in the passive scalar field, \red{additional high-order harmonic components are then mixed to fit the difference between the physical structure and the mother wavelets, which then biases the extracted scaling \citep{Huang2011PRE}}. In this aspect MSA can be considered as a data-driven type of WTMM without any transform. \red{Similar with the  wavelet leader, \citet{Welter2013PRE} proposed a multifractal analysis based on the amplitude extrema of intrinsic mode functions, which can be retrieved by the empirical mode decomposition algorithm \citep{Huang1998EMD}. In some synthesized data tests, a proposed parameter $m=2$ is used to determine the search domain for the local amplitude maxima.}

\section{Case verification and applications}
\subsection{Fractional Brownian motion }

\begin{figure}[htb]
\centering
\includegraphics[width=0.65\linewidth,clip]{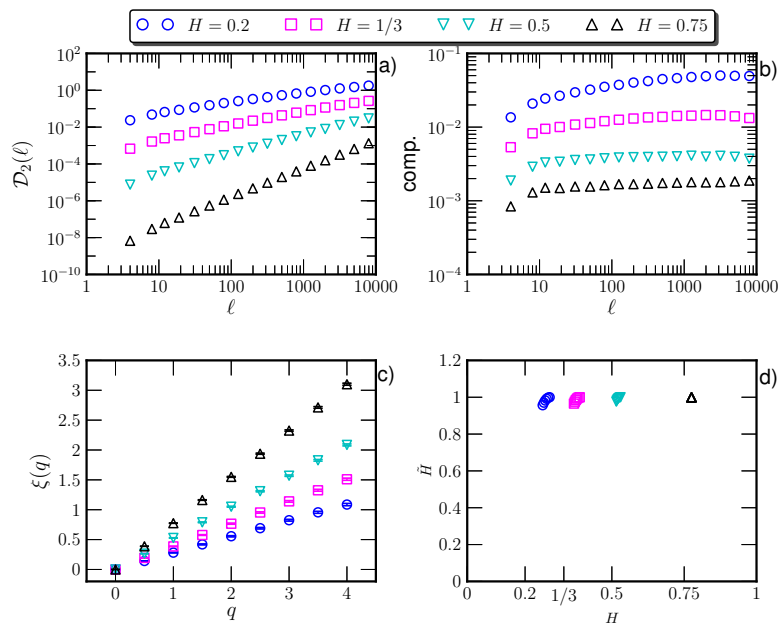}
\caption{ (Color online) a) Calculated second-order $\mathcal{D}_2(\ell)$ by MSA for fractional
Brownian motion with Hurst number $H=0.2$ ($\ocircle$), $H=1/3$ ($\square$),
$H=0.5$ ($\bigtriangledown$) and $H=0.75$ ($\triangle$), respectively. Power-law
behavior can be observed for all $H$. b) Compensated curves $\mathcal{D}_2(\ell)\times \ell ^{-2H}$. For display clarity, these curves have been vertical
shifted. c) Experimental scaling exponents $\xi_H(q)$ in the range $0\le
q\le4$. The errorbar implies the standard deviation from 100 realizations.
d) Measured singularity spectrum $f(\alpha)$.
% The inset shows the estimated
%Hurst number $\tilde{H}$ for each given $H$, compared with the theoretical %results by the solid line.
 }\label{fig:fBmMSA}
\end{figure}

\begin{figure}[htb]
\centering
\includegraphics[width=0.65\linewidth,clip]{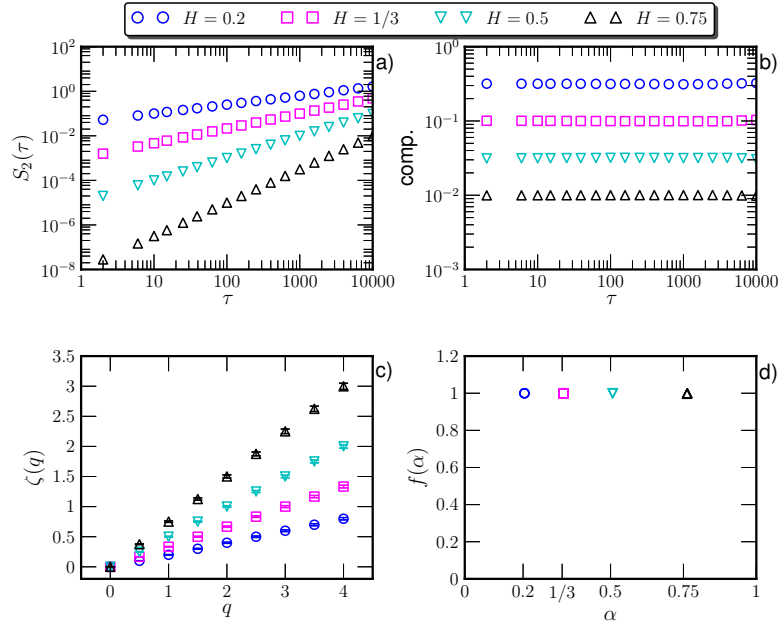}
\caption{ (Color online) Experimental results for the SF analysis: a) the measured second-order SFs $S_2(\tau)$; b) compensated curves $S_2(\tau)\times\tau^{-2H}$; c) measured scaling exponent $\zeta(q)$, and d) the measured singularity spectrum $f(\alpha)$ versus $\alpha$. Symbols are the same as in figure \ref{fig:fBmMSA}. }\label{fig:fBmSF}
\end{figure}

\begin{figure}[htb]
\centering
\includegraphics[width=0.65\linewidth,clip]{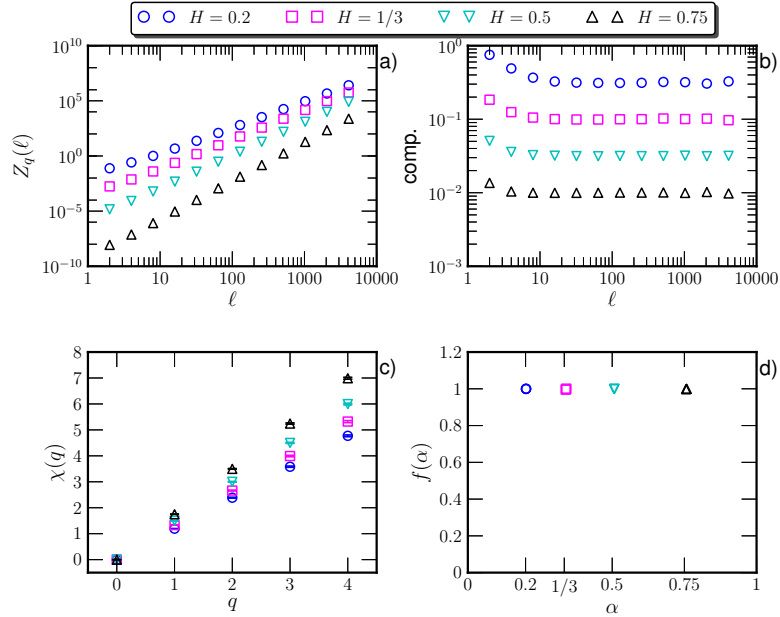}
\caption{ (Color online)  Experimental results for the wavelet leader: a) the
measured second-order  $Z_2(\ell)$; b) compensated curves $Z_2(\ell)\times\tau^{-(2H+2)}$;
c) measured scaling exponent $\chi(q)$, and d) the measured singularity
spectrum $f(\alpha)$ versus $\alpha$. The symbols are the same as in figure \ref{fig:fBmMSA}. }\label{fig:fBmWL}
\end{figure}

\begin{figure}[htb]
\centering
\includegraphics[width=0.5\linewidth,clip]{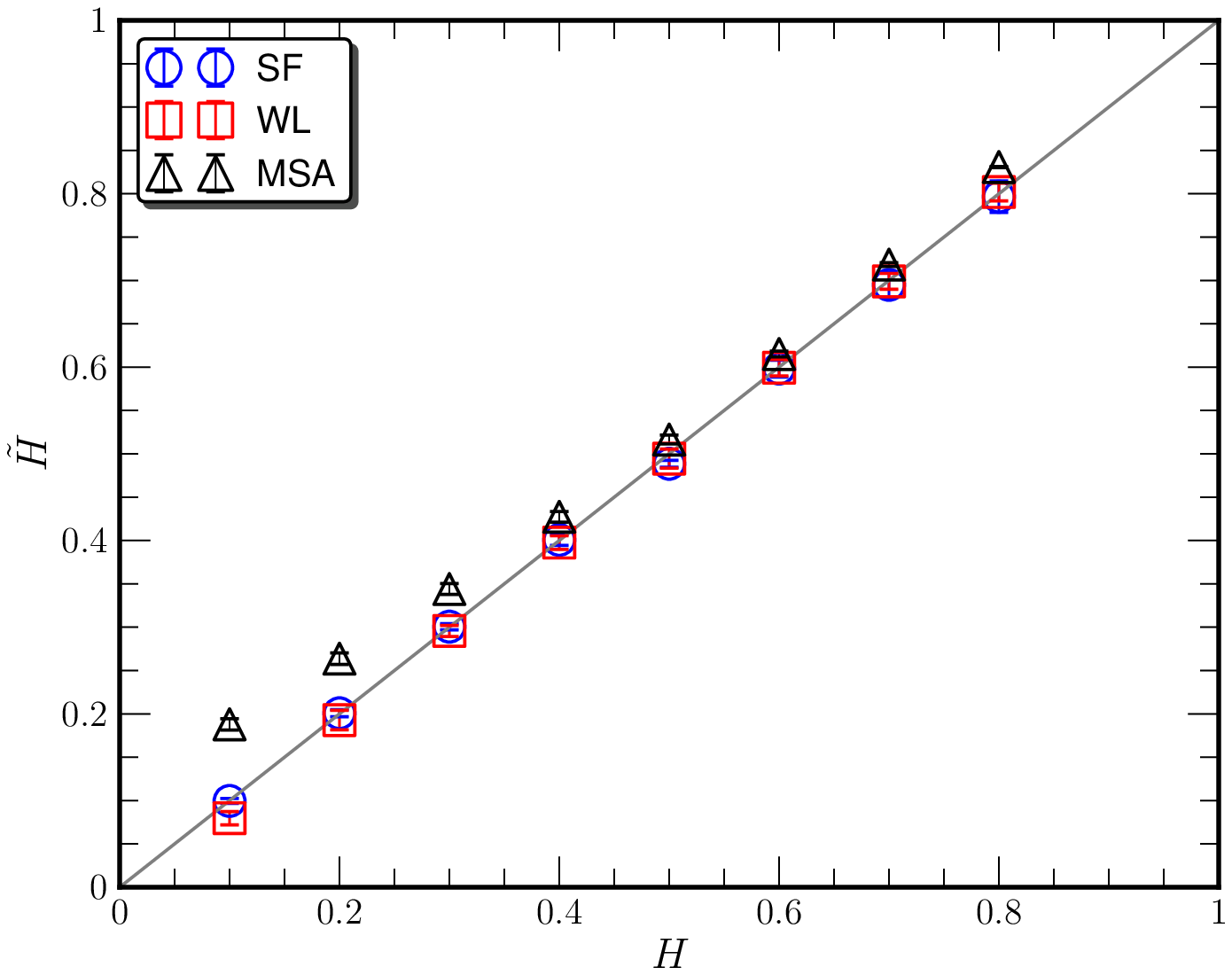}
\caption{ (Color online) Comparison of the measured Hurst number $\tilde{H}$ versus the given $H$. The theoretical value is indicated by the the solid line. }\label{fig:fBm}
\end{figure}

Fractional Brownian motion (fBm) is a generalization of the classical Brownian motion. It was introduced by~\citet{Kolmogorov1940} and extensively studied by Mandelbrot and co-workers in the 1960s \citep{Mandelbrot1968}. Since then, it is considered as a classical scaling stochastic process in many fields \citep{Beran1994,Rogers1997,Doukhan2003}. In the multifractal context, fBm is a simple self-similar process. More precisely, the measured SF scaling exponent $\zeta_H(q)$ is linear with the moment order $q$, i.e.,
$\zeta_H(q)=qH$,
in which  $H\in (0,1)$ is the Hurst number. The above linear scaling relation has been verified by various methodologies, such as, classical SF, wavelet-based method, detrended fluctuation analysis and detrended structure-function, to list a few. In the present work, a fast Fourier transform based Wood-Chan algorithm \citep{Wood1994} is used to synthesize the fBm data with 100 realizations, and $10^6$ points for each $H$.

Figure \ref{fig:fBmMSA}\,a) shows respectively the calculated second-order $\mathcal{D}_2(\ell)$ by MSA for Hurst number $H=0.2$ ($\ocircle$), $H=1/3$ ($\square$), $H=0.5$ ($\bigtriangledown$), and $H=0.75$ ($\triangle$). The power-law behavior is observed for all $H$ considered here. \red{To emphasize the agreement between the measured $\xi_H(q)$ and the theoretical value, compensated curves of $\mathcal{D}_2(\ell)\ell^{-2H}$ are shown in figure \,\ref{fig:fBmMSA}\,b).} For display clarity, these curves have been vertical shifted. Visually, when $H<1/3$, the measured $\xi(q)$ deviates from the theoretical prediction. Such outcome may be related to MSA itself or the fBm date generation algorithm. Further investigation will be conducted hereof. Figure \ref{fig:fBmMSA}\,c) shows the measured $\xi(q)$, \red{which are estimated in the range $10<\ell<10,000$ by least-square-fitting. The errorbars indicate the standard deviation from 100 realizations (same in the following)}. These curves demonstrate for all the cases the linear dependence of the measured scaling exponent $\xi(q)$ with $q$, whose slope is $\tilde{H}$. In other words, multifractality can successfully be detected by MSA for all $H$ values (including $H<1/3$). Consider the so-called singularity spectrum, which is
defined through a Legendre transform as,
\begin{equation}
\alpha=\frac{\upd\xi(q)}{\upd q},\quad f(\alpha)=\min_{q}\left\{ \alpha q -\xi(q)+1 \right\}.
\end{equation}
For a monofractal process, $\alpha$ is independent with $q$, e.g., $\alpha=H$, and $f(\alpha)=1$
\citep{Frisch1995Book}. In practice, for a prescribed $q$, a wider range
of $\alpha$ has, a more intermittent the process is. Figure \ref{fig:fBmMSA}\,d)
shows the measured $f(\alpha)$ in the range $0\le q \le4$.  It confirms the
monofractal property of the fBm process.

\red{Figure \ref{fig:fBmSF} lists the results from SF: a) the measured second-order SFs, $S_2(\tau)$, b) the compensated curves $S_2(\tau)\times \tau^{-2H}$, c) the measured scaling exponents $\zeta(q)$  and d) the corresponding singularity spectrum $f(\alpha)$ versus $\alpha$, respectively. Visually, a plateau is observed for all $\tau$, showing a perfect agreement between the detected scaling and theory. Here the scaling exponents $\zeta(q)$ are also estimated in the same the range $10<\tau<10,000$ by least-square-fitting. It shows that $\zeta(q)$ is linear with $q$, and the measured singularity spectrum $f(\alpha)$ detects correctly the monofractal property of the fBm process.}

\red{Figure \ref{fig:fBmWL} shows the results from the second-order WLs. Power-law behavior is observed for all $H$. Compensated curves $Z_2(\ell)\times \ell^{-(2H+2)}$ show a clear plateau when $\ell\ge 10$. Similar as MSA, misalignment is observed when $\ell<10$, which may be due to the fBm generator used in this study. Scaling exponents are then estimated in $10<\tau<10,000$. The measured $\chi(q)$ and the corresponding singularity spectrum $f(\alpha)$ capture the monofractal property of the fBm. Note that the singularity spectrum of WLs is defined as \citep{Huang2011PRE},
\begin{equation}
\alpha=\frac{\upd \chi(q)}{\upd q}-1, f(\alpha)=\min_{q}\left\{\alpha q-\chi(q)+1+q \right\}.
\end{equation} }

\red{Figure \ref{fig:fBm} shows the comparison of the different estimated Hurst numbers $\tilde{H}$, which are calculated by linear fitting of the measured scaling exponent $\xi(q)$, $\zeta(q)$ and $\chi(q)$, respectively. Visually, SF and WL provide almost the same performance, while MSA slightly overestimates $H$ when $H>1/3$. All methods considered here confirm the monofractal property of the fBm process.
Considering that the turbulent data are much different from the simple fBm, as already shown in Ref. \cite{Huang2011PRE}, both SF and WL are strongly influenced by the real turbulent structures.}

We provide a comment here on the deviation of the measured $\xi(q)$ from the theoretical prediction when $H\le 1/3$. In the context of extreme point based MSA, the intrinsic structure of fBm is presented by these extrema. Therefore how $\xi(q)$ will behave is determined by the relation between the Hurst number $H$ and the distribution of the extreme points, as well as the fBm data generation.  \red{Or in other words, the retrieved $\xi(q)$ relies on  the dynamic behavior of the process itself, which is deeply related with the distribution of the extrema point.} This is beyond the topic of this paper. We will present more details on this topic in the follow-up studies.

\subsection{Lagrangian velocity}
\begin{figure}[htb]
\centering
\includegraphics[width=0.85\linewidth,clip]{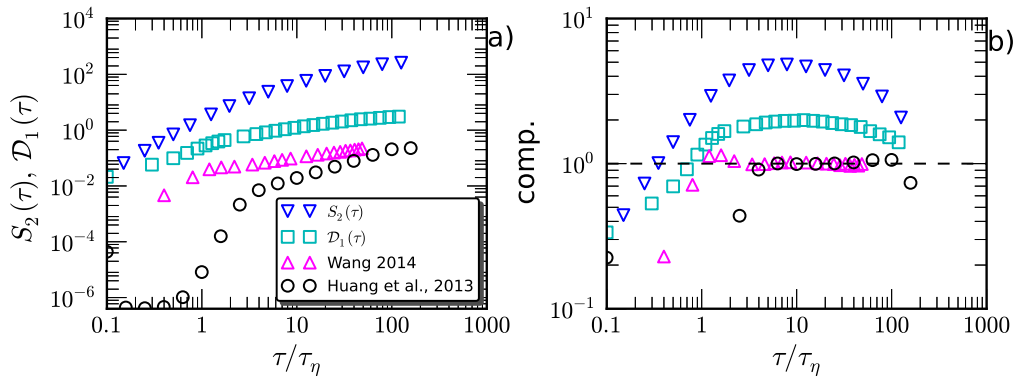}
\caption{(Color online) a) Results from various methods: classical SF
($\bigtriangledown$) and the Hilbert-based method ($\ocircle$) for the
second-order statistics, the trajectory-segment method ($\triangle$) and MSA $\mathcal{D}_1(\tau)$ ($\square$) for the first-order statistics. For the Hilbert-based statistics, frequency has been converted to time by $\tau=1/\omega$. b) The corresponding curves compensated by the dimensional scalings, i.e. $S_2(\tau)\sim \varepsilon\tau$ and $S_1(\tau)\sim (\varepsilon\tau)^{1/2}$. There is no plateau from SF, in consistency with other reports in the literature. The following convincing scaling ranges can be observed: about $2<\tau/\tau_{\eta}<60$ from MSA, $2<\tau/\tau_{\eta}<50$ from the trajectory segment analysis and $10<\tau/\tau_{\eta}<100$ from the Hilbert-based method, which verify the prediction of the Kolmogorov-Landau's phenomenological theory. For display clarity, these curves have been vertical shifted.} \label{fig:Lagrangian}
\end{figure}

The Lagrangian velocity SF has been extensively studied. Because the time scale separation in the Lagrangian frame is more Reynolds number dependent than the length scale case in the Eulerian frame, the finite Reynolds number influence becomes stronger, making the Lagrangian velocity scaling relation quite controversial \citep{Falkovich2012PoF}. More specifically, this is recognized as a consequence of mixing between large-scale structures and energetic small-scale structures, e.g., vortex trapping events \citep{Toschi2005JOT,Huang2013PRE}. The $q$th-order SF of the velocity component $u_{i}\ (i=1,2,$ or $3)$ is defined as:
\begin{equation}
S_{q}(\tau)\equiv \langle[u_{i}(t+\tau)-u_{i}(t)]^{q}\rangle,
\label{sf}\end{equation}
where $\tau$ is an arbitrary time separation scale. From dimensional analysis, the 2nd-order SF is supposed to satisfy~\citep{Falkovich2012PoF}:
\begin{equation}
S_{2}(\tau)=C_{0} \varepsilon \tau,
\label{2sf}\end{equation}
where $\varepsilon$ is the rate of energy dissipation per unit mass and $C_{0}$ is assumed as a universal constant at high Reynolds numbers.

To analyze this problem, we adopted the data from direct numerical simulation (DNS), implemented for the isotropic turbulence in a $2048^{3}$ cubic domain. The boundary conditions are periodic along each spatial direction and kinetic energy is continuously provided at few lowest wave number components. A fine resolution of $dx\sim\eta$ (the Kolmogorov scale) ensures resolving the detailed small-scale velocity dynamics. The Taylor scale $\lambda$ based Reynolds number $Re_\lambda$ is about $400$. Totally $0.2$ million Lagrangian particle samples are collected, each of which has about one integral time life span. During the evolution process, the velocity and velocity derivatives are recorded at each $\tau_\eta/20$, which $\tau_\eta$ is the Kolmogorov time. More numerical details can be found in Ref.~\cite{Benzi2009PRE} and references therein.

Recently, this database has been analyzed respectively by \citet{Huang2013PRE}, and \citet{Wang2014PoF} to identify the inertial scaling behavior. The former study employed the Hilbert-based approach, in which different scale events are separated by the empirical mode decomposition without any \textit{a priori} basis assumption and the corresponding frequency $\omega$ is extracted by the Hilbert spectral analysis. They observed clearly an inertial range of $0.01<\omega\tau_{\eta}<0.1$, i.e. $10<\tau/\tau_{\eta}<100$. The scaling exponents $\zeta(q)$ agree well with the multifractal model (see details in Ref. \cite{Huang2013PRE}). The latter one studies the extrema of the fluid particle acceleration, which physically can be considered as the boundary markers between different flow regions. With the help of the so-called trajectory segment structure, the clear scaling range does appear. Because of interpolation inaccuracy (noise), DNS data need to be particularly smoothed~\citep{Wang2014PoF}, which may lead to some artificial input.

Figure \ref{fig:Lagrangian}\,a) shows the numerical results from various methods: classical SF ($\bigtriangledown$) and the Hilbert-based method ($\ocircle$) for the second-order statistics, the trajectory-segment method ($\triangle$) and MSA ($\square$) for the first-order statistics. Except for SF, clear power-law behaviors can be observed for others. To emphasize this, curves compensated by the dimensional scalings, i.e. $S_2(\tau)\sim \varepsilon\tau$ and $S_1(\tau)\sim (\varepsilon\tau)^{1/2}$, are plotted in Fig.\,\ref{fig:Lagrangian}\,b). The SF curve does not show any plateau, which is consistent with reports in the literature \citep{Falkovich2012PoF,Sawford2011PoF}, even for high $Re$ cases as $Re_{\lambda}\simeq 815$ experimentally or $Re_{\lambda}\simeq 1000$ numerically. The absence of the clear inertial range makes the Kolmogorov-Landau's phenomenological theory quite controversial. In comparison, the MSA curve shows a convincing plateau in the range $2<\tau/\tau_{\eta}<60$.

It need to mention that for this DNS database the inertial range has been recognized as $10<\tau/\tau_{\eta}<100$ for both the single particle statistics using the Hilbert-based methodology \citep{Huang2013PRE} and the energy dissipation statistics to check the Lagrangian version refined similarity hypothesis  \citep{Huang2014JFM}. Here the inertial range $2<\tau/\tau_{\eta}<60$ detected by MSA is due to the different scale definition.

% Figure~\ref{fig:Lagrangian} (a) shows the results from SF for the velocity magnitude $|u|$ and the three components $u_1$, $u_2$ and $u_3$, which fail to see any scaling in the inertial range. A recent progress by \citet{Wang2014PoF} studies the extrema of the fluid particle acceleration, which physically can be considered as the boundary markers between different flow regions. The so-called trajectory segment is the part bounded by adjacent markers along Lagrangian paths. Because of interpolation inaccuracy (noise), DNS data need to be particularly smoothed~\citep{Wang2014PoF}, which may lead to some artificial input.
%
%
% Following the similar idea, but we apply here MSA to search the local acceleration extrema at different $s$ levels. Generally the data noise will induce spurious extremal points with small segment lengths. When $s$ increases, such induced extrema disappear and the influence from noise on the segment statistics can be neglected. In this sense MSA is robust to resist the noise influence and can be regarded as a general extension (at least in one-dimensional space). The results from MSA and trajectory segment~\citep{Wang2014PoF} are shown in Fig.~\ref{Lagrangian} (b). A convincing scaling range about $2<\Delta\tau/\tau_{\eta}< 60$ can be observed. The Ref.~\citep{Wang2014PoF} plot also verifies convincingly the prediction by Eq.~\ref{2sf}. Relatively the scale of trajectory segments is much confined.

\subsection{2D turbulence velocity field}

 \begin{figure}[htb]
 \centering
\includegraphics[width=0.55\linewidth,clip]{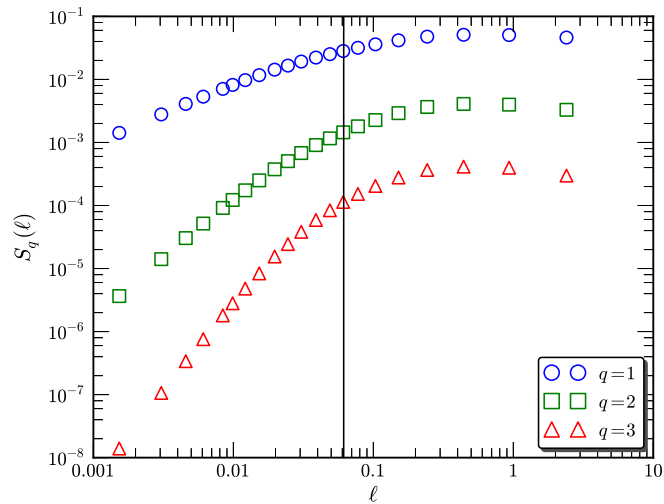}
 \caption{(Color online) Measured structure-function $S_q(\ell)$ for the two-dimensional
turbulent velocity. Due to the scale-mixing in SF analysis, there is no power-law behavior.} \label{fig:2DSF}
 \end{figure}

\begin{figure}[htb]
\centering
\includegraphics[width=0.95\linewidth,clip]{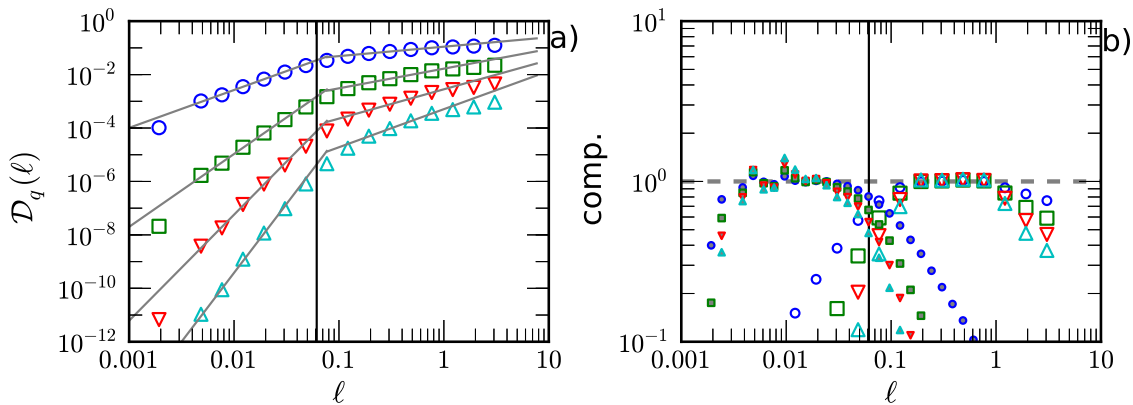}
\caption{(Color online) a) Measured $q$th-order $\mathcal{D}_q(\ell)$ from MSA. A dual-cascade
power-law behavior is observed in the range $0.004<\ell<0.04$ for the forward
enstrophy cascade, and $0.15<\ell<1$ for the inverse energy cascade. b) The curves
compensated by the least square fitted scaling exponent $\zeta(q)$ to emphasize the power-law behavior. The vertical solid line indicates the forcing scale  $\ell_f=0.0614.$} \label{fig:2DMSA}
\end{figure}

\begin{figure}[htb]
\centering
\includegraphics[width=0.85\linewidth,clip]{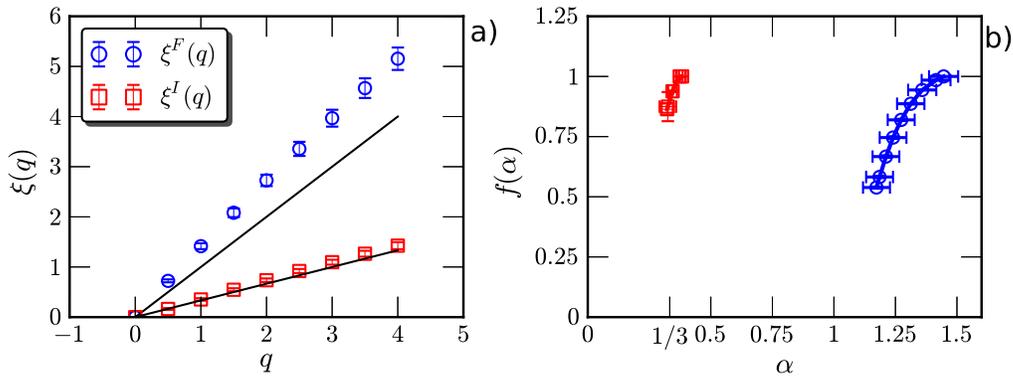}
\caption{(Color online) a) Fitted scaling exponents $\xi(q)$ for both the forward
enstrophy  ($\ocircle$) and inverse energy ($\square$) cascades. The two solid
lines represent respectively the Krainchan's prediction $\xi(q)=q$ for
the forward cascade and $\xi(q)=q/3$ for the inverse cascade. b) Measured
singularity spectrum $f(\alpha)$. The inset shows the enlargement part in $0.2<\alpha<0.5$. A broader range of $1.17\le\alpha\le1.45$ and $0.54\le
f(\alpha)\le1$ for the forward enstrophy cascade indicates a stronger intermittency.} \label{fig:2DExponents}
\end{figure}

Two-dimensional (2D) turbulence is an ideal model for the large scale movement of the ocean or atmosphere \citep{Boffetta2012ARFM,Kraichnan1980RPP,Tabeling2002PhysRep,Bouchet2012PhysRep}. We recall here briefly the main theoretical results of 2D turbulence advocated by \citet{Kraichnan1967PoF}.

The 2D Ekman-Navier-Stokes equation can be written as
\begin{equation}
\partial_t \mathbf{u}+\mathbf{u}\cdot\nabla\mathbf{u} =-\nabla p+ \nu \bigtriangledown^2 \mathbf{u}-\alpha \mathbf{u} +\mathbf{f}_{u},  \label{eq:NSvorticity}
\end{equation}
in which $\mathbf{u}(\mathbf{x},t)=[u(\mathbf{x},t),v(\mathbf{x},t)]$ is the velocity vector, $\nu$ is the fluid viscosity, $\alpha$ is the Ekman friction and $\mathbf{f}_{u}$ is an external source of energy inputting into the whole system at scale $\ell_{\mathbf{f}}=1/k_{\mathbf{f}}$~\citep{Boffetta2007JFM}.
Parallelly, the vorticity $\omega=\nabla\times\mathbf{u}$ equation is
\begin{equation}
\partial_t\omega++\mathbf{u}\cdot\nabla\omega = \nu
\bigtriangledown^2 \omega-\alpha \omega +\mathbf{f}_{\omega}.\label{eq:ENS}
\end{equation}
To keep the whole system balance, two conservation laws emerge. The first one is the so-called energy conservation, inducing an inverse energy cascade from the forcing scale $\ell_{\mathbf{f}}$ to large scales, i.e., $\ell>\ell_{\mathbf{f}}$, which then leads to a five-third law in the Fourier space above the forcing scale, i.e.,
\begin{equation}
E_u(k)=C \epsilon^{2/3}_{\alpha}  k^{-5/3},\quad k_{\alpha}\ll k\ll k_\mathbf{f} \textrm{ for the inverse energy cascade},
\label{eq:Energy}\end{equation}
where $E_u(k)$ is the Fourier power spectrum of $\mathbf{u}$,  $\epsilon_{\alpha}$ is the energy dissipation by the Ekman friction, $k_{\alpha}$ is the Ekman friction scale and $k_{\mathbf{f}}=1/\ell_{\mathbf{f}}$ is the forcing scale. Below the forcing scale $\ell<\ell_{\mathbf{f}}$, the enstrophy conservation law yields
\begin{equation}
E_u(k)=C' \eta^{2/3}_{\nu}  k^{-3},\quad k_{\mathbf{f}}\ll k\ll k_{\nu}
\textrm{ for the forward enstrophy cascade},\label{eq:Enstrophy}
\end{equation}
in which $\eta_{\nu}$ is the enstrophy dissipation by the viscosity $\nu$ and $k_{\nu}$ is the viscosity scale. This double-cascade 2D turbulence theory has been recognized as one of the most important results in turbulence since Kolmogorov's 1941 work \citep{Falkovich2006PhysToday}.

For the last few decades, numerous experiments and numerical simulations have been devoted to verify the above mentioned forward and inverse cascades \citep{Bruneau2005PRE,Rutgers1998PRL,Kellay1998PRL,Bernard2006NatPhys,Boffetta2007JFM,Falkovich2011PRE,Tan2014PoF} with partially verification of the theory by Kraichnan. For example, \citet{Boffetta2010PRE} performed a very high resolution numerical simulation, up to a grid number $N=32,768^2$. They stated that due to the scale separation problem, numerically the dual-cascade requires a very high resolution for verification.
The inverse and forward cascades were observed for the third-order velocity structure function as predicted by the theory~\cite[see Figure 3]{Boffetta2010PRE}. For the inverse cascade, it is found that the inverse energy cascade is almost nonintermittent~\citep{Nam2000PRL,Tan2014PoF}, while for the forward enstrophy cascade, the intermittency effect still remains as a open question. This is because, according to the convergency condition, the structure-function requires a Fourier spectrum $E(k)\sim k^{-\beta}$ with $\beta\in(1,3)$,  the $\beta$-limitation \citep{Huang2010PRE}.
Coincidentally, the Fourier scaling exponent for the enstrophy cascade is $\beta\ge 3$, see equation~(\ref{eq:Enstrophy}) and discussion in Ref.~\cite{Boffetta2010PRE}. Theoretically, \citet{Nam2000PRL} found that the Ekman friction leads to an intermittent forward enstrophy cascade~\citep{Bernard2000EPL}, which has been verified indirectly by studying the passive scalar field, instead of the vorticity field \citep{Boffetta2002PRE}. More recently, this claim has been confirmed by~\citet{Tan2014PoF} using Hilbert spectral analysis. A log-Poisson model without justice is proposed to fit the forward enstrophy cascade scaling exponent, see detail in Ref.~\cite{Tan2014PoF}. However, to identify whether the forward enstrophy cascade is intermittent or not is still a challenge in the sense of data analysis.

The DNS data for present analysis is based on a fully resolved vorticity field simulation \citep{Boffetta2007JFM}, with an artificially added friction coefficient. Numerical integral of equation (\ref{eq:ENS}) is performed by a pseudo-spectral, fully dealiased on a doubly periodic square domain of size $L=2\pi$ with $N^2=8192^2$ grid points~\citep{Boffetta2007JFM}. The key parameters adopted are $\nu=2\times10^{-6}$, $\alpha=0.025$ and the energy input wave number $k_{\mathbf{f}}=100$ with very short correlation in time. The velocity field is be solved from the Poisson equation of the stream function $\psi$, i.e.,  $\mathbf{u}=[\partial_y \psi,-\partial_x \psi]$. Totally, five snapshots with $8192^2\times 5=3.36\times10^8$ data points are used for analysis. More details of this database can be found in Ref.~\cite{Boffetta2007JFM}.

The conventional SFs are shown in figure\,\ref{fig:2DSF}  and no clear scaling range can be observed, neither any indication of the aforementioned two regimes. As discussed above and also in Ref.~\cite{Tan2014PoF}, such outcome can be ascribed to the scaling mixing in SF analysis \citep{Huang2010PRE}. For comparison the results from MSA are shown in figure\,\ref{fig:2DMSA}\,a), in which two regimes with different scaling relations appear, specifically in the range $0.004<\ell<0.04$ for the forward enstrophy cascade and $0.15<\ell<1$ for the inverse energy cascade. To emphasize the observed power-law behavior, the compensated curves are then displayed in figure\,\ref{fig:2DMSA}\,b) by using the fitted scaling exponents $\xi(q)$.  Two clear plateaus appear, confirming the existence of the  dual-cascade process in the 2D turbulence. The scaling exponents $\xi(q)$ are then estimated in these scaling ranges by a least-square-fitting algorithm. Figure \ref{fig:2DExponents}\,a) shows the measured dual-cascade $\xi(q)$. The theoretical predictions by equations (\ref{eq:Energy}) (i.e. $q/3$) and (\ref{eq:Enstrophy}) (i.e. $q$) are indicated by solid lines. Note that the measured forward cascade curve is larger than the theoretical one. A similar observation for the Fourier power spectrum has been reported in Ref.~\cite{Boffetta2007JFM}, which is considered as an influence of the fluid viscosity $\nu$.

To detect the multifractality, the singularity spectrum $f(\alpha)$ is then estimated, which is shown in figure\,\ref{fig:2DExponents}\,b). Based on the fBm case test, one can conclude that the inverse cascade is nonintermittent as expected; while the forward enstrophy cascade shows a clear sign of multifractality: a broad change of $\alpha$ and $f(\alpha)$.
\red{Note that $\alpha$ is the generalized Hurst number. The value range $\alpha\in (1.17,1.45)$ implies a Hurst number $H>1$, much larger than the one indicated by the equation~(\ref{eq:Enstrophy}), i.e., $H=1$. Such outcome may be an effect of the logarithmic correction \citep{Pasquero2002PRE}, while it is reported by \cite{Tan2014PoF} that the logarithmic correlation for the vorticity field is weak.} It also has to point out here that the measured $\alpha$ and $f(\alpha)$ could be a function of $\nu$ or the Ekman friction \citep{Boffetta2007JFM,Boffetta2012ARFM,Tan2014PoF}. Systematic analysis of the 2D velocity field with different parameter is necessary in the future for deeper insights.

\section{Discussion and Conclusions}

\red{In summary, we propose in this paper a multiscale statistical method, namely multi-level segment analysis, without employing any decomposition. The statistical properties of extremal points at multiple $s$ level inherit the intrinsic multi-scale physics of complex systems. For each $s$, the corresponding extremal point series defines the so-called segment, which can be characterized by the separation distance and the function difference at adjacent extrema. As an important extension of the existing work, the MSA method introduced in this paper proves meaningful in complex data analysis. This new approach has been verified by revealing the monofractal property of the synthesized fBm processes, while the Hurst number $H$ is slightly overestimated especially when $H<1/3$. When applied to two numerical turbulent datasets, i.e. the high-resolution Lagrangian 2D turbulence, MSA shows interesting outcome. The conventional methodologies, such as SF, fail to detect the inertial scaling of the velocity field, which is now recognized partially as the $\beta$-limitation, and partially as contamination by the energetic structures \citep{Huang2013PRE}. Very differently, MSA detects successfully the clear scaling ranges for both cases. More precisely for 2D turbulence, the retrieved multifractal property of the forward enstrophy confirms the theoretical prediction by \cite{Nam2000PRL}. More systematic study of such multifractality and parametric dependance (e.g. the fluid viscosity) is important in future to provide a better understanding of 2D turbulence.}

% Some preliminary properties of MSA and cast tests have been discussed, %including the standard fractional Brownian motion, Lagrangian and 2D turbulence.
Finally, we provide the following general remarks:
\begin{itemize}
\item[1] In principle, MSA is generally applicable without special requirements on the data itself, such as periodicity, the Fourier spectrum slope $\beta$ ( steeper than $3$), noise perturbation and unsmoothness structures.
\item[2] The length scale in the context of MSA is determined by the functional structure rather than being an independent input. At different $s$ levels the segments are different and thus the scales as well, which conforms with the multi-scale physics.
\item[3] A serious deficiency of the conventional SF is the strong mixing of different correlation and scaling regimes due to sample averaging, i.e. the filtering (as infrared and ultraviolet) effect. To define the segment structure helps to annihilate such filtering and extract the possible scaling relations in the respective scale regimes, which has been proved from analyzing the Lagrangian and 2D turbulence data.
\item[4] \red{Moreover irregular sampling (e.g. missed points) is a common problem for data collection as in the geophysical experiments, LDV (Laser Doppler Velocimetry) measurement, etc. However, typically methods as Fourier analysis and others require uniform spacial data points. Such trouble is easy to overcome in MSA since only extremal instead of all the functional points are involved.}

\end{itemize}

\ack
\red{The authors are thankful for the comments and suggestions given by the two anonymous reviewers.} LW acknowledges the funding support by national science foundation China (NSFC) under
the grant Nos. 11172175 and 91441116. This research work by YH is partially sponsored by the NSFC under Grant Nos. 11202122 and 11332006. \red{We thank Professor P. Abry from Laboratoire de Physique, CNRS, and ENS, Lyon (France) for providing his wavelet leader codes.} We thank Professor G. Boffetta and Professor F. Toschi for providing us the  DNS data, which are freely available from the iCFDdatabase: {http://cfd.cineca.it} for public DNS database of the 2D turbulence and the Lagrangian turbulence.

%\bibliographystyle{jphysicsB}
%\bibliography{all}% Produce

\end{document}